\documentclass[a4paper]{jpconf}
\usepackage{graphicx}
\usepackage{subfigure}
\usepackage{amsfonts}
\usepackage{amsmath}
\usepackage{amssymb}
\newcommand{\T}{\mathbb{T}}
\begin{document}
\title{Robust matching rules for real quasicrystals}

\author{Pavel Kalugin$^1$ and Andr\'e Katz$^2$}

\address{$^1$Laboratoire de Physique des Solides, CNRS, Universit\'e Paris-Sud, Universit\'e Paris-Saclay, F-91405 Orsay, France.}
\address{$^2$Directeur de recherche honoraire, CNRS, France}
\ead{kalugin@lps.u-psud.fr}

\begin{abstract}
We consider the problem of extraction and validation of matching rules, directly from the phased diffraction data of a quasicrystal, and propose an algorithmic procedure to produce the rules of the shortest possible range. We have developed a geometric framework to express such rules together with a homological mechanism enforcing the long-range quasiperiodic order. This mechanism tolerates the presence of defects in a robust way.
\end{abstract}
\section{Introduction}
It is commonly acknowledged that the long-range order in quasicrystals depends on a hypothetical order propagation mechanism usually referred to as {\em matching rules}. However, so far, this understanding has been applied to the structure determination on a case by case basis only, for instance by trying to interpret the observed structure as a decoration of a tiling already known to have matching rules. In \cite{kalugin2018}, we suggested that the exploration of matching rules should instead be the primary goal in solving quasicrystalline structures. This article is a short presentation of this program; for further details and bibliography see \cite{kalugin2018} and references therein.
\section{Homology-based matching rules}
By the very nature of our program, the model should be locally derivable from the atomic structure only. This requirement leads naturally to modeling matching rules in terms of simplicial tilings with vertices located at the atomic positions and labeled by their local environment. Still, even a locally deterministic triangulation may yield ambiguous results in degenerate cases (e.g. there exist two ways to cut a square in two triangles). This justifies the use of homological methods, since they allow for construction of matching rules inherently insensitive to such artificial ambiguities.
\par
Let us recall some definitions and notations of \cite{kalugin2018}. We encode the matching rules of a simplicial tiling in a geometrical object obtained by gluing together all prototiles by their matching faces. This yields a finite cellular complex $B$ equipped with the metric data inherited from the $d\mbox{-dimensional}$ physical space $E$. In \cite{kalugin2018} we call $B$ a flat-branched semisimplicial complex (or {\em FBS-complex}). We assume that the tiling of $E$ can be lifted to a corrugated $d\mbox{-surface}$ in a larger space $E\oplus F$ of dimension $n>d$, which can be seen as a graph of the ``phason coordinate function'' $\varphi: E \to F$. This graph is a subset of a periodic pattern in $E\oplus F$. Factoring the entire construction over the translations of the corresponding lattice $\mathcal{L} \subset E\oplus F$ results in wrapping the lifted tiling over $\T^n=(E\oplus F)/\mathcal{L}$. This wrapping can be pulled back to the FBS-complex $B$ via the lifting map $\beta: B \to \T^n$.
\par
\begin{figure}%
	\centering
	\subfigure[]{\label{fig:x-y}\includegraphics[width=10pc]{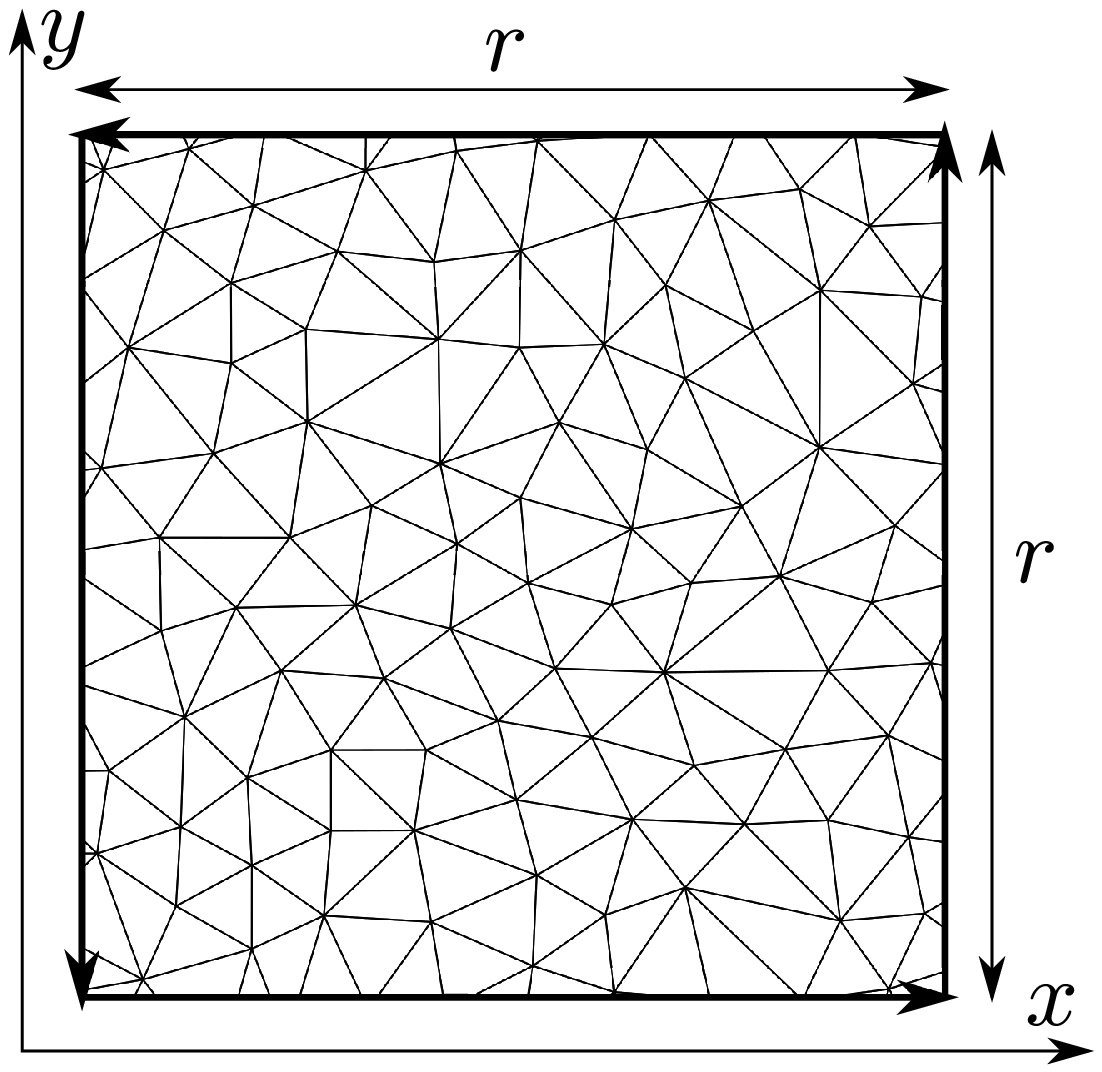}}
	\subfigure[]{\label{fig:x-y-phi}\includegraphics[width=13pc]{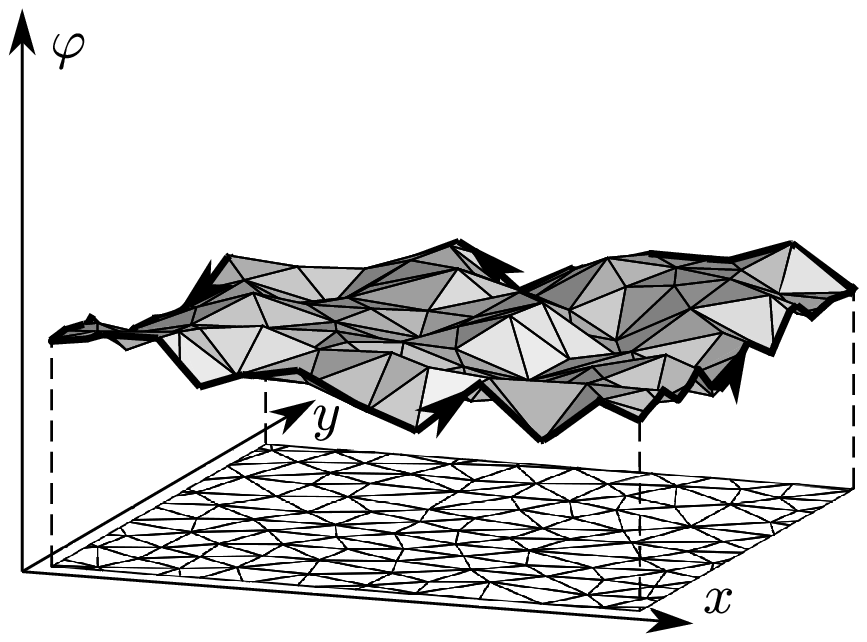}}
	\subfigure[]{\label{fig:x-phi}\includegraphics[width=13pc]{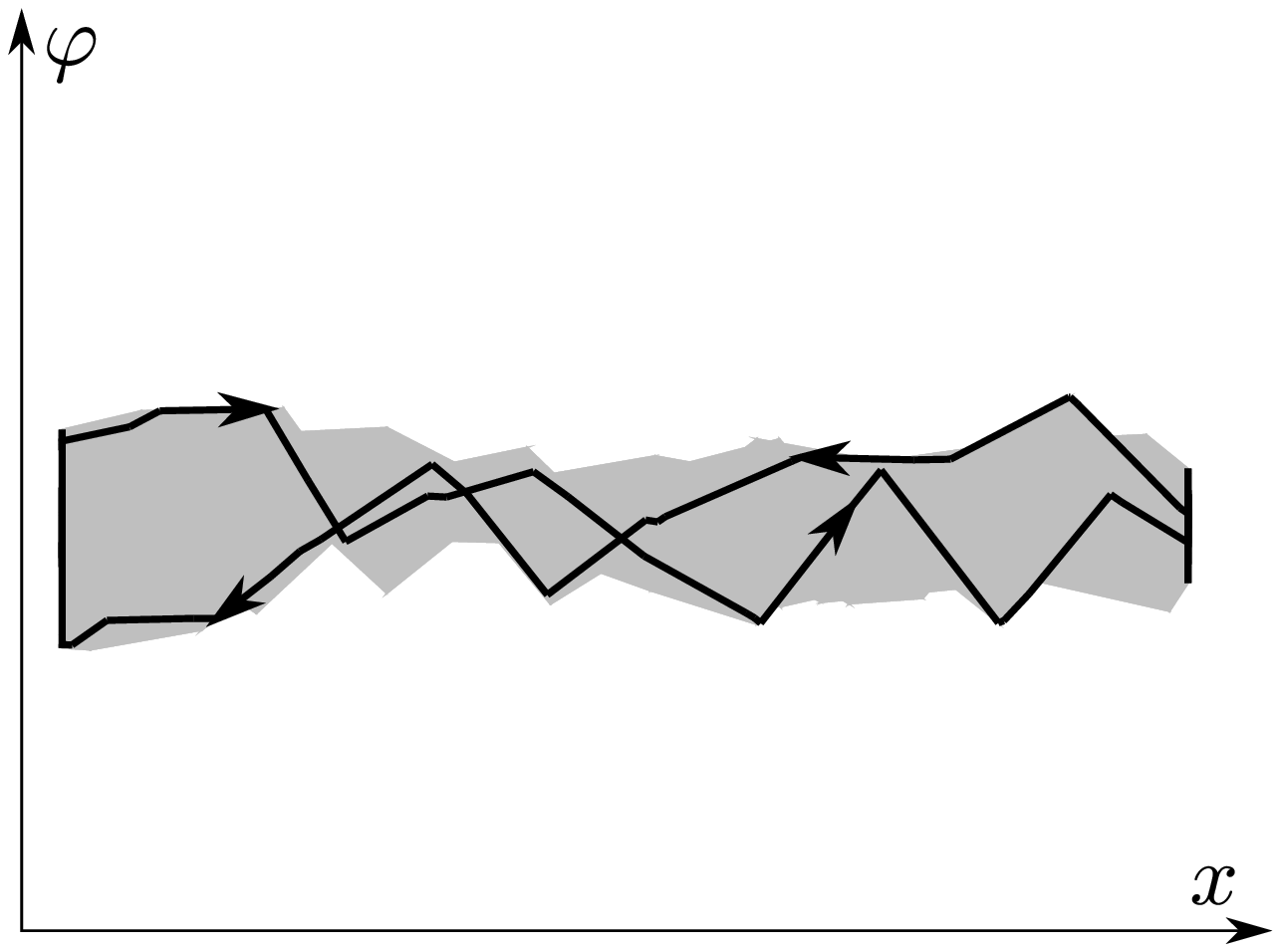}}%
	\caption{Three views of a square patch of a simplicial tiling for $d=2$: \subref{fig:x-y} represents the original patch, \subref{fig:x-y-phi} depicts the lifted tiling with only one dimension of the phason coordinate $\varphi$ shown and \subref{fig:x-phi} represents the projection of the patch on the plane $(x, \varphi)$ (only the shadow of the lifted patch and the projection of its boundary are shown). The oriented area of the lifted patch projected onto the plane $(x, \varphi)$ equals the integral of the mixed form $\omega=d\varphi \wedge dx$ over the lifted patch. If $\beta: B \to \T^n$ is slope locking, this integral evaluates to a boundary term and its absolute value is bounded by $Kr$ for some $K>0$. Therefore, the difference of the value of $\varphi$ averaged over the edges of the square parallel to the $x$ axis is bounded by $K$.}
	\label{fig:triptique}
\end{figure}
To illustrate the key idea of the homology-based matching rules, let us assume for the moment that the phason gradient is constant (that is $\varphi$ is affine). Since the phason gradient depends on $d(n-d)$ real parameters, one needs the same number of conditions to fix it. In particular, the condition of zero phason gradient is equivalent to the annihilation of a certain $d(n-d)\mbox{-dimensional}$ space $T$ of $d\mbox{-forms}$ on $E\oplus F$ by the graph of $\varphi$. This space (we shall refer to its elements as {\em mixed forms}) is spanned by products of constant $1\mbox{-forms}$ in $F$ and $(d-1)\mbox{-forms}$ in $E$. 
\par
\begin{figure}[h]
	\begin{minipage}[b]{18pc}
	\includegraphics[width=18pc]{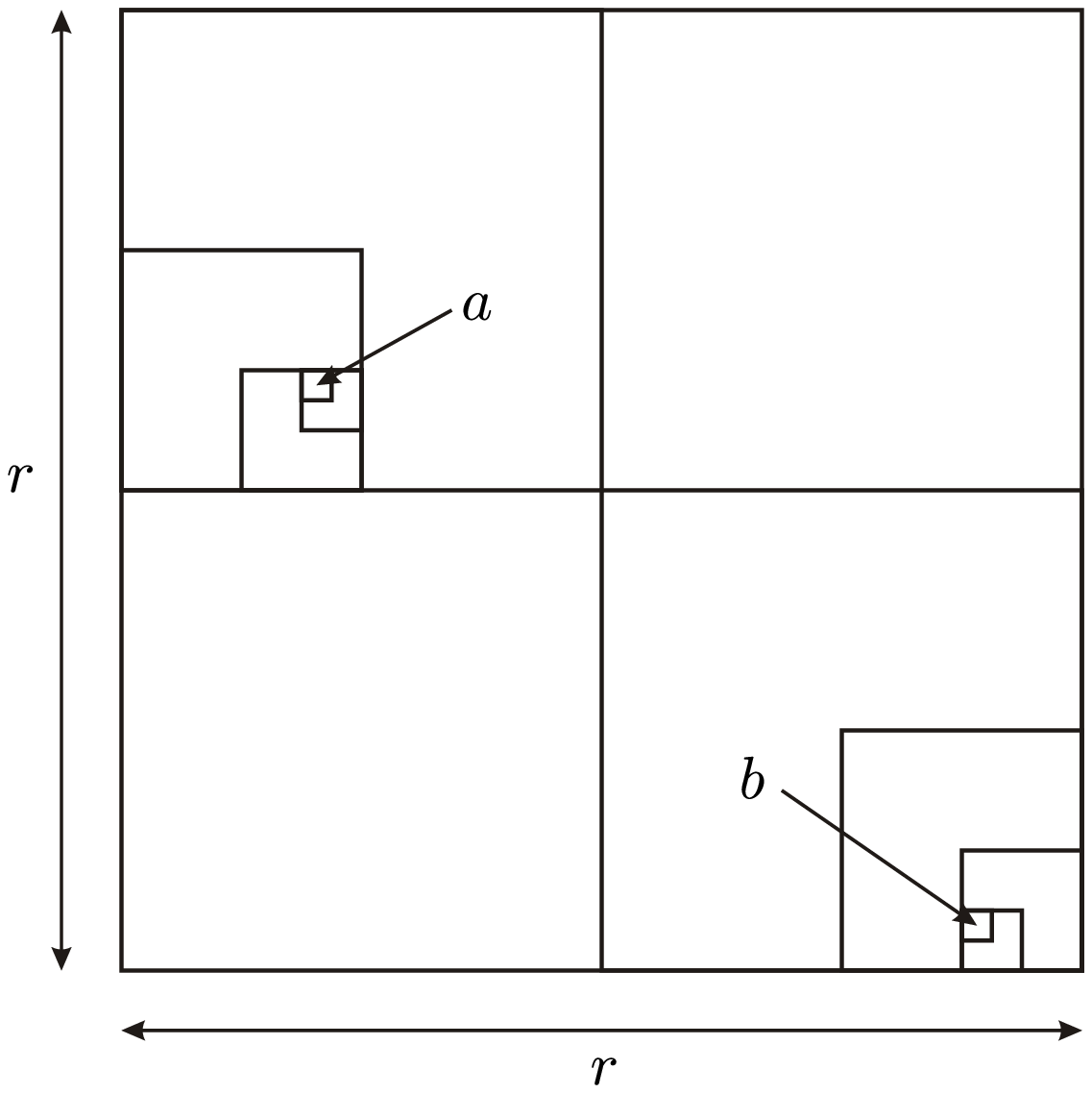}
	\caption{\label{fig:hierarchy} A $d\mbox{-dimensional}$ cube of edge length $r$ and two sequences of nested cubes obtained by repeated partitioning and converging to the points $a$ and $b$ (five iterations are shown).}
	\end{minipage}\hspace{1.5pc}
    \begin{minipage}[b]{18pc}
    \includegraphics[width=18pc]{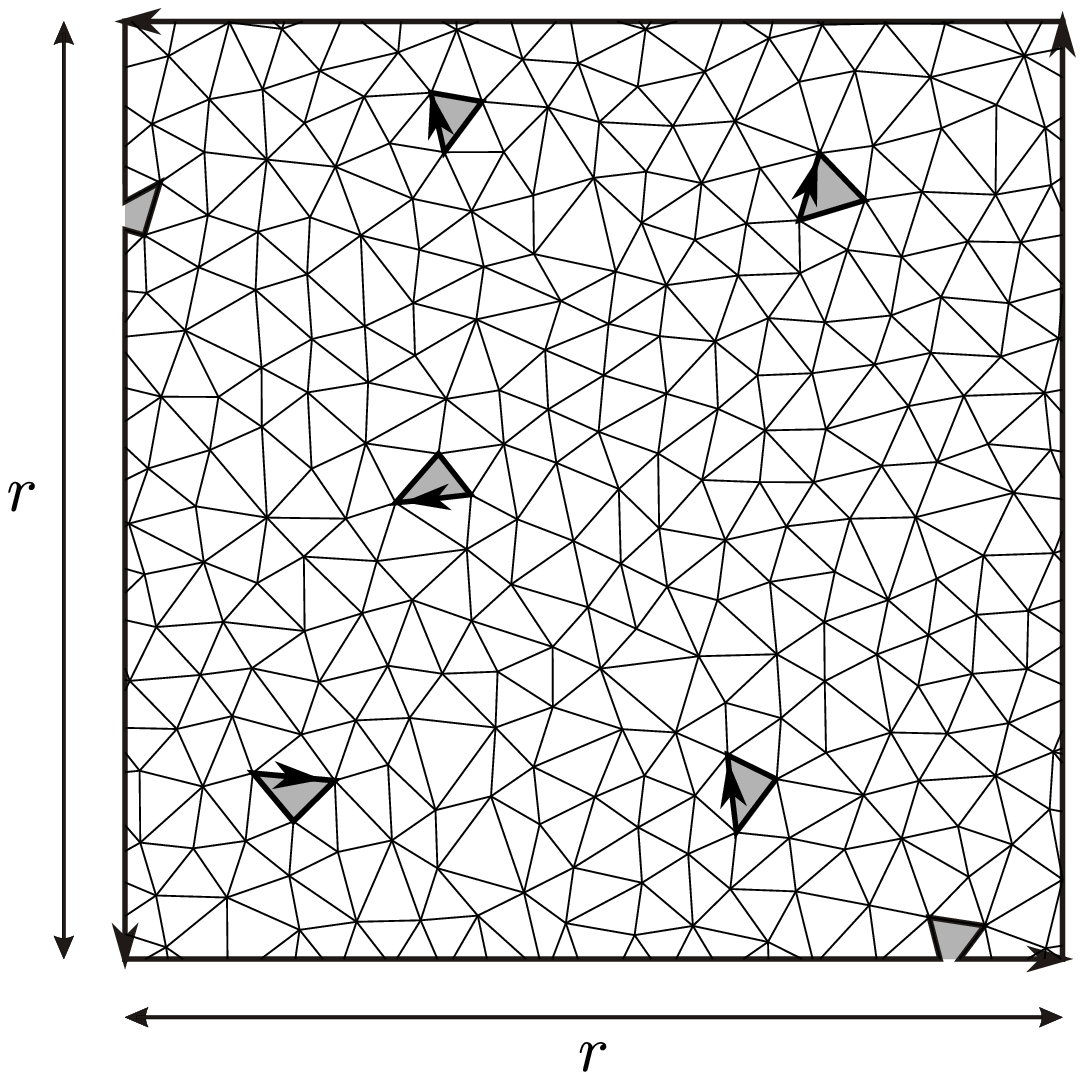}
    \caption{\label{fig:defects}A square patch of a simplicial tiling with defects. The thick line represents the oriented boundary of its defect-free part. Its length scales with $r$ as $\mathcal{O}(\max(\varepsilon r^d, r^{d-1}))$, where $\varepsilon$ is the density of defects.}
    \end{minipage}
\end{figure}

For the actual tiling models, individual lifted tiles do not generally annihilate $T$, and the matching rules imply only an asymptotic annihilation of $T$ by large patches of the lifted tiling. We shall thus require such annihilation for the image of every $d\mbox{-cycle}$ on $B$ under the lifting map $\beta$ (for brevity we do not make distinction between constant forms on the space $E\oplus F$ and those on its factor $\T^n$). If $\beta$ has such property (called {\em slope locking} in \cite{kalugin2018}), then for any mixed form $\omega \in T$, its pullback on $B$ is a coboundary and the integral of $\omega$ over the graph of $\varphi$ is reduced to a boundary term. As illustrated by Figure \ref{fig:triptique}, in this case the difference between the values of $\varphi$ averaged over the opposite faces of an arbitrarily oriented cube of edge length $r$ in $E$ is bounded by some constant $K$ independent on $r$. Therefore the difference between the values of $\varphi$ averaged over any two cubes sharing a common face is also bounded by $K$. By partitioning a cube of edge length $r$ into $2^n$ cubes of edge length $r/2$ (see Figure \ref{fig:hierarchy}) and applying the same argument for each dimension, we obtain the upper bound $Kd$ for the difference between values of $\varphi$ averaged over the original cube and that averaged over any of the cubes of the partition. This procedure can be iterated down to the scale of individual tiles. Since the number of iterations grows as $\log_2(r)$ as $r \to \infty$, one has $\left\|\varphi(a)-	\varphi(b)\right\| <Kd\log_2(r) + \mathrm{const}$, yielding
\begin{equation}
\label{logrule}
\|\varphi(x)\|=\mathcal{O}(\log(\|x\|))
\end{equation}
and thus fixing the slope of the lifted tiling.
\par
Let us show now that the homology-based matching rules are robust with respect to the presence of defects, which means that a small concentration of defects results in a small overall phason gradient. Defects can be conveniently introduced by assuming that the FBS-complex $B$ of the perfect structure (the one for which the lifting map $\beta$ is slope locking) is contained in a larger FBS-complex $\check B \supset B$, such that the extension of $\beta$ to $\check B$ may not be slope locking. The defects thus correspond to the simplices of the complement $\check B \backslash B$ (the corresponding tiles are shown shaded on Figure \ref{fig:defects}). Let $\varepsilon$ stand for the concentration of the defects. Then, following the reasoning above, we obtain that the difference of $\varphi$ averaged over the opposite faces of a cube of edge length $r$ is bounded by $K_1 + \varepsilon K_2 r$ for some positive real $K_1$ and $K_2$, and the maximal value of the phason gradient is limited by a term proportional to the density of defects:
$$
\|\varphi(x)\|=\mathcal{O} (\max(\log(\|x\|), \varepsilon \|x\|).
$$
\section{Working with real quasicrystals}
\begin{figure}[h]
	\includegraphics[width=20pc]{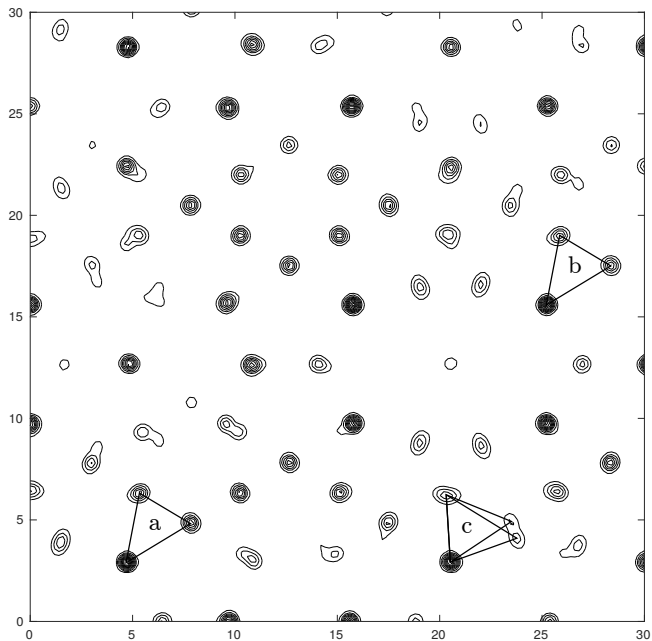}\hspace{1.5pc}%
	\begin{minipage}[b]{16pc}\caption{\label{fig:cut2}The contour plot of the Fourier synthesis of the phased density for the icosahedral quasicrystal $\mathrm{Cd}_{5.7}\mathrm{Yb}$ (obtained from the diffraction data of \cite{takakura2007atomic}), restricted to a square $30\text{\AA}\times 30 \text{\AA}$ in the two-fold symmetry plane. Triangles (a) and (b) are $\mathcal{L}\mbox{-equivalent}$. The dumbbell-shaped peak near the label (c) arises from cutting of two different atomic surfaces. The two maxima of this peak are too close to each other for both to be retained in $\mathcal{T}_0$. In such situations, the retained maximum is chosen at random, and this choice may be different for $\mathcal{L}\mbox{-equivalent}$ peaks. Thus, the FBS-complex $B_0$ will include 2-simplices corresponding to both of the triangles (c).}
	\end{minipage}
\end{figure}

The exploration of matching rules starts with the phased diffraction data, which is seen not as an approximation to the actual structure but as a source of information about the local environments. We start by assigning to each atomic surface an anchoring point in $\T^n$. Then we set up the vertices of the future tiling in the physical space, placing them roughly at the  at the peaks in the phased density (some peaks should be skipped to avoid too short distances between vertices, see Figure \ref{fig:cut2}). The vertices are also adjusted to the projections of the anchoring points of the corresponding atomic surfaces. Applying Delaunay triangulation to the vertex set yields a simplicial tiling $\mathcal{T}_0$ of finite local complexity (note that in practice we work with a finite patch of $\mathcal{T}_0$). The vertices of $\mathcal{T}_0$ are characterized by an atomic surface label and a translation of the lattice $\mathcal{L}$. We call two simplices having all their vertices related by the same translation $\mathcal{L}\mbox{-equivalent}$. Factoring $\mathcal{T}_0$ over the $\mathcal{L}\mbox{-equivalence}$ yields an FBS-complex $B_0$ together with the lifting map $\beta_0: B_0 \to \T^n$ (the latter takes each vertex of $B_0$ to the anchoring point of the corresponding atomic surface).
\par
In the (unlikely) case of $\beta_0$ being slope locking, our goal is achieved. Otherwise we have to proceed further with reduction and refinement of $B_0$. The reduction consists in gradual elimination of simplexes of $B_0$, starting with those having vertices located at the peaks of density of uncertain shape or those occurring rarely in $\mathcal{T}_0$, until the restriction of $\beta_0$ on the remaining sub-complex $B \subset B_0$ becomes slope-locking. It may occur that the slope locking is not achieved before too many simplices are eliminated and $B$ does not allow for tiling of the entire space (this is the case, for instance, when $B$ does not admit a real cycle which image in $\T^n$ corresponds to the winding of $E$ over $\T^n$). In this case, $B_0$ has to be refined and then the reduction start over again. The refinement consists in enriching the vertex labels by the labels of neighboring vertices in $\mathcal{T}_0$. This procedure can be repeated, each time encoding into $B_0$ the information about larger local configurations. Naturally, the first time the slope locking conditions are achieved would correspond to the matching rules of the shortest possible range.
\par
Suppose now that the proposed algorithm yielded an FBS-complex $B$ with a slope-locking lifting map $\beta: B \to \T^n$. To produce an actual structure model, the matching rules encoded by $B$ require a validation. An ultimate way to validate the matching rules would consist in constructing a tiling obeying them and comparing the predicted diffraction intensities with the experimental data. This is an intricate problem, since there is no systematic way to construct a tiling satisfying a given set of matching constraints. It might be therefore interesting before trying this to perform a partial validation, by checking other predictions of the model, namely the value of the total atomic density and the distribution of the density between different atomic surfaces (see \cite{kalugin2018}). Finally, even if the existence of an infinite tiling free of matching defects remains unproven, the robust nature of the proposed matching rules makes tilings with a small concentration of defects a valid structure model as well.
\section*{References}
\bibliographystyle{iopart-num}
\bibliography{robust_matching_icq14}
\end{document}